# Stellar coronal astronomy

Fabio Favata (fabio.favata@rssd.esa.int)
*Astrophysics Division of ESA's Research and Science Support Department – P.O. Box 299, 2200 AG Noordwijk, The Netherlands*

Giuseppina Micela (giusi@oapa.astropa.unipa.it)
*INAF – Osservatorio Astronomico G. S. Vaiana – Piazza Parlamento 1, 90134 Palermo, Italy*

**Abstract.**
Coronal astronomy is by now a fairly mature discipline, with a quarter century having gone by since the detection of the first stellar X-ray coronal source (Capella), and having benefitted from a series of major orbiting observing facilities. Several observational characteristics of coronal X-ray and EUV emission have been solidly established through extensive observations, and are by now common, almost text-book, knowledge. At the same time the implications of coronal astronomy for broader astrophysical questions (e.g. Galactic structure, stellar formation, stellar structure, etc.) have become appreciated. The interpretation of stellar coronal properties is however still often open to debate, and will need qualitatively new observational data to book further progress. In the present review we try to recapitulate our view on the status of the field at the beginning of a new era, in which the high sensitivity and the high spectral resolution provided by *Chandra* and XMM-*Newton* will address new questions which were not accessible before.

*Full paper available at*
ftp://astro.esa.int/pub/ffavata/Papers/ssr-preprint.pdf